\documentclass[aps,pra]{revtex4}
\usepackage{graphicx}

\begin{document}

\title{Multiparty Quantum Secret Sharing
\thanks{Email: zhangzj@wipm.ac.cn}}

\author{Zhan-jun Zhang$^{1,2}$, Yong Li$^3$ and Zhong-xiao Man$^2$ \\
{\normalsize $^1$ School of Physics \& Material Science, Anhui University, Hefei 230039, China} \\
{\normalsize $^2$ Wuhan Institute of Physics and Mathematics,
Chinese Academy of Sciences, Wuhan 430071, China} \\
{\normalsize $^3$ Department of Physics, Huazhong Normal
University, Wuhan 430079, China} \\ {\normalsize Email:
zhangzj@wipm.ac.cn}}

\date{\today}
\maketitle

\begin{minipage}{420pt}
Based on a quantum secure direct communication (QSDC) protocol
[Phys. Rev. A69(04)052319], we propose a $(n,n)$-threshold scheme
of multiparty quantum secret sharing of classical messages (QSSCM)
using only single photons. We take advantage of this multiparty
QSSCM scheme to establish a scheme of multiparty secret sharing of
quantum information (SSQI), in which only all quantum information
receivers collaborate can the original qubit be reconstructed. A
general idea is also proposed
for constructing multiparty SSQI schemes from any QSSCM scheme. \\

\end{minipage}\\

Suppose Alice wants to send a secret message to two distant
parties, Bob and Charlie. One of them, Bob or Charlie, is not
entirely trusted by Alice. And she knows that if the two guys
coexist, the honest one will keep the dishonest one from doing any
damages. Instead of giving the total secret messages to any one of
them, it may be desirable for Alice to split the secret messages
into two encrypted parts and send each one a part so that no one
alone is sufficient to obtain the whole original information but
they collaborate. To gain this end classical cryptography can use
a technique called as secret sharing [1,2], where secret messages
are distributed among $n$ users in such a way that only by
combining their pieces of information can the $n$ users recover
the secret messages. Usually this kind of protocols are divided
into classes, where from the $n$ receivers, $m$ can collaborate to
produce the desired result. In this paper, we will focus on a
$(n,n)$ scheme, where all the receivers need to collaborate to
obtain the desired message.

Recently this concept has been generalized to a quantum scenario
[3]. The quantum secret sharing (QSS) is likely to play a key role
in protecting secret quantum information, e.g., in secure
operations of distributed quantum computation, sharing
difficult-to-construct ancilla states and joint sharing of quantum
money [6], and so on. Hence, after the pioneering QSS work
proposed by using three-particle and four-particle GHZ states [3],
this kind of works on QSS attracted a great deal of attentions in
both theoretical and experimental aspects [4-16].  All these works
[3-16] can be divided into two kinds, one only deals with the QSS
of classical messages (i.e., bits)[5-6,8-11,13-14], or only deals
with the QSS of quantum information [4,7,12,15-16] where the
secret is an arbitrary unknown state in a qubit; and the other [3]
studies both, that is, deals with QSS of classical messages and
QSS of quantum information simultaneously. In all those schemes
[3,5-6,8-11,13-14] dealing with the QSS of classical messages
(bits), entangled states are used with only an exception [13]
where multi-particle product states are employed. On the other
hand, in all those schemes [3,4,7,12,15,16] dealing with the QSS
of quantum information, multi-particle entangled states are used.

Recently, a particular quantum secure direct communication (QSDC)
protocol has been proposed by Deng and Long [17], in which only
single photon state is used. In this paper, based on Deng-Long's
QSDC protocol, we propose a scheme of multiparty quantum secret
sharing of classical messages (QSSCM) by using only single
photons. And then we take advantage of this multiparty QSSCM
scheme to establish a scheme of multiparty secret sharing of
quantum information (SSQI), where the secret is an arbitrary
unknown quantum state in a qubit. We will show that multi-particle
entangled states are unnecessary in our multiparty SSQI scheme.
Finally, we will propose a general idea for constructing
multiparty SSQI schemes from any QSSCM scheme.

Now let us turn to our multiparty QSSCM scheme. For convenience,
let us first describe a three-party QSSCM scheme. Suppose Alice
wants to send a secret message to two distant parties, Bob and
Charlie. One of them, Bob or Charlie, is not entirely trusted by
Alice, and she knows that if the two guys coexist, the honest one
will keep the dishonest one from doing any damages. The two
receivers, Bob and Charlie, can infer the secret message only by
their mutual assistance. Our following three-party QSSCM scheme
can achieve this goal with 5 steps.

(a) Bob prepares a batch of $N$ single photons randomly in one of
four polarization states $|H\rangle=|0\rangle$,
$|V\rangle=|1\rangle$,
$|u\rangle=1/\sqrt{2}(|0\rangle+|1\rangle)$,
$|d\rangle=1/\sqrt{2}(|0\rangle-|1\rangle)$.  For convenience,
\{$|H\rangle$,$|V\rangle$\} is refereed to as the rectilinear
basis and \{$|u\rangle$,$|d\rangle$\} the diagonal basis
hereafter. Then he sends this batch of photons to Charlie.

(b) After receiving these photons, for each photon Charlie
randomly choose a unitary operation from $I$, $U$ and $U_H$ and
performs this unitary operation on it. Here
$I=|0\rangle\langle0|+|1\rangle\langle1|$ is an identity operator,
$U=|0\rangle\langle1|-|1\rangle\langle0|$ and
$U_H=(|0\rangle\langle0|+|1\rangle\langle0|+|0\rangle\langle1|-|1\rangle\langle1|)/\sqrt{2}$
is a Hadamard gate operator. The nice feature of the $U$ operation
is that it flips the state in both measuring bases, i.e.,
$U|0\rangle=-|1\rangle$, $U|1\rangle=|0\rangle$,
$U|u\rangle=|d\rangle$, $U|d\rangle=-|u\rangle$. And the nice
feature of $H$ is that it can realize the transformation between
the rectilinear basis and the diagonal basis, i.e.,
$U_H|H\rangle=|u\rangle$, $U_H|V\rangle=|d\rangle$,
$U_H|u\rangle=|H\rangle$, $U_H|d\rangle=|V\rangle$. After his
encryptions, he sends the photons to Alice. The purpose of
choosing a set of three unitary operations is to protect the
channel between Alice and Charlie from Bob's interception. For
example, if Charlie chooses randomly a unitary operation from only
$I$ and $U$, Bob could intercept the channel between Alice and
Charlie. He already has all the information about the state of the
photon, and can readily check what is the transformation Alice
did, so then he can retrieve the complete message without the help
of Charlie.

(c) Alice stores most of the single photons and selects randomly a
subset of single photons. Alice publicly announces the position of
the selected photons. For each selected photon Alice randomly
selects one action from the following two choices. One is that
Alice lets Bob first tell her the initial state of the photon and
then lets Charlie tell her which unitary operation he has
performed on it; the other is that Charlie first tells Alice which
unitary operation he has performed on the photon and then Bob
tells Alice the initial state of the photon. Alice's strategy of
choosing two actions is to prevent either Bob's or Charlie's
intercept-resend attacks. Then Alice first performs the same
unitary operation as Charlie has performed on the photon and then
measures the photon by using the basis the initial state belongs
to. After her measurements, Alice can determine the error rate. If
the error rate exceeds the threshold, the process is aborted.
Otherwise, the process continues and Alice performs unitary
operations (either $I$ or $U$) on the stored photons to encode her
secret messages. That is, if Alice wants to encode a bit '0', she
performs the identity unitary operation $I$; if Alice wants to
encode a bit '1', she performs the unitary operation
$U=|0\rangle\langle1|-|1\rangle\langle0|$.  Alice sends these
encoded photons to Charlie.

(d) After Charlie receives these encoded photons, if Bob and
Charlie collaborate, both Bob and Charlie can obtain Alice's
secret message by using correct measuring basis for each encoded
photon. On the other hand, if Bob and Charlie do not collaborate,
then both Bob and Charlie can not get access to Alice's secret
message with 100\% certainty.

(e) Alice publicly announces a small part of her secret messages
for Bob and Charlie to check whether the photons travelling from
Alice site to Charlie's site have been attacked, which is called
message authentification. If the photons are attacked, the
eavesdropper Eve can not get access to any useful information but
interrupt the transmissions.

So far we have proposed the three-party QSSCM scheme based on Deng
and Long's QSDC protocol [17] by using single photons. The
security of the present three-party QSSCM scheme is the same as
the security of Deng and Long's QSDC protocol [17], that is, it
depends completely on the step when Charlie sends the photon batch
to Alice. As proven in [17], the scheme is also unconditionally
secure. Incidentally, one can easily find that if Alice sends the
encoded photons to Bob instead of Charlie, then the resultant
scheme works securely also.

Now let us generalize the three-party QSSCM scheme to a $n$-party
($n\geq 4)$ QSSCM scheme. Suppose that Alice is the message sender
who would like to send a massage to Bob, Charlie, Dick, \dots, and
Zach (there is totally $n$ receivers). The first step of the
$n$-party ($n\geq 4)$ QSSCM scheme is the same as that in the
three-party QSSCM scheme. For completeness, this step is also
included as follows.

(I) Bob prepares a batch of $N$ single photons randomly in one of
four polarization states $|H\rangle=|0\rangle$,
$|V\rangle=|1\rangle$,
$|u\rangle=1/\sqrt{2}(|0\rangle+|1\rangle)$,
$|d\rangle=1/\sqrt{2}(|0\rangle-|1\rangle)$. Then he sends this
batch of photons to Charlie.

(II) After receiving these photons, for each photon Charlie
randomly chooses a unitary operation from $I$, $U$ and $U_H$ and
performs this unitary operation on it. After his encryptions, he
sends the encrypted photons to the next receiver, say, Dick. Dick
encrypts randomly the encoded photons in the same way as Charlie,
then he sends the photons to the next receiver, and so on. Similar
procedure is repeated until Zach finishes his encryptions. After
Zach's encryptions, he sends the encrypted photons to Alice.

(III) Alice stores most of the single photons and selects randomly
a subset of single photons. And then Alice publicly announces the
position of the selected photons. To prevent any receiver's
intercept-resend attack, for each selected photon, Alice randomly
selects a receiver one by one and let him or her tell her this
receiver's message till she obtains all receivers's messages. Here
Bob's message is the initial state of the photon, while Charlie's,
(Dick's, \dots, Zeck 's) message is which unitary operation he has
performed on the photon. Alice performs in turn the same unitary
operations as Zeck's, the $(n-2)$th receiver's, \dots, Dick's and
Charlie's unitary operations on the photon and then measures this
photon by using the basis the initial state belongs to. After her
measurements, Alice can determine the error rate. If the error
rate exceeds the threshold, the process is aborted. Otherwise, the
process continues and Alice performs unitary operations (either
$I$ or $U$) on the stored photons to encode her secret messages.
Alice sends these encoded photons to Zach.

(IV) After Zach receives these encoded photons, if Zach and the
other $n-1$ receivers (Bob, Charlie, Dick, \dots, the $(n-1)$th
receiver) collaborate, they can obtain Alice's secret message by
using correct measuring basis for each encoded photon. On the
other hand, if all the receivers do not collaborate, then none of
them can get access to Alice's secret message with 100\%
certainty.

(V) Alice publicly announces a small part of her secret messages
for all the receivers to check whether the photons travelling from
Alice site to Zach's site have been attacked, which is called
message authentification. If the photons are attacked, the
eavesdropper Eve can not get access to any useful information but
interrupt the transmissions.

So far we have established a $n$-party QSSCM scheme by using
single photons. The security of the present $n$-party QSSCM scheme
is the same as the security of three-party QSSCM scheme, which is
also unconditionally secure. Incidentally, as mentioned
previously, one can easily find that if Alice sends the encoded
photons to any other receiver instead of Zach, then the resultant
scheme works securely also.

Now let us move to propose a multiparty SSQI scheme. Before this,
let us briefly review the secure teleportation of an unknown
quantum state [19]. Suppose that Alice wants to send to Bob an
unknown state $\alpha |H\rangle_u +\beta|V\rangle_u$ in her qubit.
Bob prepares a photon pair in any Bell state, say,
$|\Phi^+\rangle_{ht} =\frac{1}{\sqrt{2}}(|H\rangle_{h}
|H\rangle_{t} +|V\rangle_{h} |V\rangle_{t}
=\frac{1}{\sqrt{2}}(|u\rangle_{h} |u\rangle_{t} +|d\rangle_{h}
|d\rangle_{t} )$. Bob sends the $t$ photon to Alice. By randomly
selecting one of the two sets of measuring basis, both Alice and
Bob can check whether the quantum channel for photon transmission
is attacked or not according to their joint actions [24]. Suppose
the quantum channel is safe and Bob successfully transits a $t$
photon to Alice. The state of the whole system can be rewritten as
\begin{eqnarray}
(\alpha |H\rangle_u +\beta|V\rangle_u)|\Phi^+\rangle_{ht} &=&
(\alpha |H\rangle_u
+\beta|V\rangle_u)\frac{1}{\sqrt{2}}(|H\rangle_{h} |H\rangle_{t}
+|V\rangle_{h} |V\rangle_{t} \nonumber \\
&=& \frac{1}{2} |\Phi^+\rangle_{ut} ((\alpha |H\rangle_h
+\beta|V\rangle_h) + \frac{1}{2} |\Psi^+\rangle_{ut} ((\alpha
|V\rangle_h +\beta|H\rangle_h)\nonumber \\
&+& \frac{1}{2} |\Phi^-\rangle_{ut} ((\alpha |H\rangle_h
-\beta|V\rangle_h) + \frac{1}{2} |\Psi^-\rangle_{ut} ((\alpha
|V\rangle_h -\beta|H\rangle_h),
\end{eqnarray}
where $|\Psi^+\rangle=(|H\rangle_{h} |V\rangle_{t} +|V\rangle_{h}
|H\rangle_{t})/\sqrt{2}$, $|\Psi^-\rangle=(|H\rangle_{h}
|V\rangle_{t} -|V\rangle_{h} |H\rangle_{t})/\sqrt{2}$ and
$|\Phi^-\rangle=(|H\rangle_{h} |H\rangle_{t}-|V\rangle_{h}
|V\rangle_{t})/\sqrt{2}$. Hence, if Alice performs a Bell-state
measurement on the two photons in her lab and tells Bob her
measurement outcome, say, $|\Phi^+\rangle$
($|\Psi^+\rangle$,$|\Psi^-\rangle$,$|\Phi^-\rangle$), then Bob can
perform a unitary operation $I=|H \rangle\langle
H|+|V\rangle\langle V|$ ($u_1=|H \rangle\langle V|+|V
\rangle\langle H|$, $u_2=|H \rangle\langle H|-|V\rangle\langle
 V|$, $u_3=|H \rangle\langle V|-|V \rangle\langle H|$) to reconstruct
the unknown state in the qubit $h$. Since the teleportation is
based on EPR pairs, so the proof of the security is the same in
essence as those in Ref.[20-24]. This is the secure teleportation
of an unknown state in a qubit. In such teleportation, Alice's
public announcement of the Bell-state measurement outcome is a
necessary step, otherwise, Bob can not reconstruct the unknown
state in his retained qubit.

Our multiparty SSQI scheme ($n\geq 3$) is almost the same as the
secure teleportation of an unknown quantum state as mentioned
above, except one point. Alice would like to send an unknown
quantum state to Bob, Charlie, Dick, \dots, and Zach (there is
totally $n$ receivers). To do this, she sends the unknown quantum
state to Bob by teleportation. But instead of public announcement
of the Bell-state measurement outcome, Alice distributes her
Bell-state measurement outcome to $n$-1 receivers without Bob by
use of the QSSCM (quantum secret sharing of classical messages)
protocol we just proposed. To reconstruct an unknown state in a
qubit, all $n$ receivers must collaborate.

In our multiparty SSQI (secret sharing of quantum information)
scheme, the multi-particle GHZ states in all other existing
multiparty SSQI schemes [3,4,7] are not necessary. Although in
[15] it is claimed that only Bell states are needed, the
identification of multi-particle GHZ state is necessary. In our
multiparty SSQI protocol, only during the teleportation step are
the use and identification of Bell states needed. In all other
steps, single photon states are enough. Hence, the present
multiparty SSQI scheme is more feasible with present-day
technique[25].

As a matter of fact, till now there have been many existing
multiparty QSSCM (quantum secret sharing of classical messages)
schemes [3,5-6,8-11,13-14]. Each of them can be combined with the
secure quantum teleportation of an unknown state to establish a
multiparty SSQI scheme. Hence the idea of combining a secure
teleportation of an unknown state with a QSSCM scheme to set up a
SSQI scheme in the present paper is a general one.

To summarize, in this paper by using single photon state instead
of entangled states (Bell states or multi-particle GHZ states) or
of multi-photon product states we have presented a multiparty
QSSCM scheme based on Deng and Long's QSDC protocol. We have also
proposed a multiparty SSQI scheme by taking advantage of our
multiparty QSSCM scheme. The idea of combining a multiparty QSSCM
scheme with the secure quantum teleportation to establish a
multiparty SSQI scheme is general.\\

\noindent {\bf Acknowledgements}

Zhang thanks to Professor Baiwen Li for his encouragement. We also
thank to the referee for his/her many constructive opinions. This
work is supported by the National Natural Science Foundation
of China under Grant No. 10304022. \\

\noindent {\bf References}

\noindent[1]  B. Schneier, Applied Cryptography (Wiley, New York,
1996) p. 70.

\noindent[2] J. Gruska,  Foundations of Computing (Thomson
Computer Press, London, 1997) p. 504.

\noindent[3] M. Hillery, V. Buzk  and A. Berthiaume, Phys. Rev. A
{\bf 59}, 1829 (1999).

\noindent[4] R. Cleve, D. Gottesman  and H. K. Lo, Phys. Rev.
Lett. {\bf 83}, 648 (1999).

\noindent[5] A. Karlsson, M. Koashi and N. Imoto, Phys. Rev. A
{\bf 59}, 162  (1999).

\noindent[6] D. Gottesman,  Phys. Rev. A {\bf 61}, 042311  (2000).

\noindent[7] S. Bandyopadhyay, Phys. Rev. A {\bf 62}, 012308
(2000).

\noindent[8] W. Tittel, H. Zbinden and N. Gisin, Phys. Rev. A {\bf
63}, 042301  (2001).

\noindent[9] V. Karimipour and A. Bahraminasab, Phys. Rev. A {\bf
65}, 042320 (2002).

\noindent[10] H. F. Chau, Phys. Rev. A {\bf 66}, 060302  (2002).

\noindent[11] S. Bagherinezhad and V. Karimipour, Phys. Rev. A
{\bf 67}, 044302  (2003).

\noindent[12] L. Y. Hsu, Phys. Rev. A {\bf 68}, 022306 (2003).

\noindent[13] G. P. Guo and G. C. Guo, Phys. Lett. A {\bf 310},
 247 (2003).

\noindent[14] L. Xiao, G. L. Long, F. G. Deng and  J. W. Pan,
Phys. Rev. A {\bf 69}, 052307 (2004).

\noindent[15] Y. M. Li, K. S. Zhang and K. C. Peng, Phys. Lett. A
{\bf 324}, 420  (2004).

\noindent[16] A. M. Lance, T. Symul, W. P. Bowen, B. C. Sanders
and P. K. Lam,  Phys. Rev. Lett. {\bf 92}, 177903  (2004).

\noindent[17] F. G. Deng and G. L. Long,  Phys. Rev. A {\bf69},
 052319 (2004).

\noindent[18]Z. X. Man, Z. J. Zhang and Y. Li, Chin. Phys. Lett.
{\bf 22}, 18 (2005).

\noindent[19] C. H. Bennett, G. Brassard C. Crepeau,  R. Jozsa, A.
Peres and W. K. Wotters, Phys. Rev. Lett. {\bf70}, 1895 (1993).

\noindent[20] C. H. Bennett, G. Brassard and N. D. Mermin, Phys.
Rev. Lett. {\bf68}, 557 (1992).

\noindent[21] H. Inamori, L. Rallan  and V. Verdral, J. Phy. A
{\bf 34}, 6913 (2001).

\noindent[22] G. L. Long and X. S. Liu, Phys. Rev. A {\bf 65},
032302 (2002) .

\noindent[23] E. Waks, A. Zeevi and Y. Yamamoto, Phys. Rev. A {\bf
65}, 052310 (2002) .

\noindent[24] F. G. Deng, G. L. Long and  X. S. Liu, Phys. Rev. A
{\bf68},  042317 (2003).

\noindent[25] Y. H. Kim, S. P. Kulik and Y. Shih, Phys. Rev. Lett.
{\bf 86}, 1370 (2001).

\enddocument